# ECAKP: ENCRYPT COLLECT AUTHENTICATE KILL PLAY

Mahmoud Mohamed El-Khouly

Department of Information Technology, Faculty of Computers and Information,
Helwan University, Cairo, Egypt
mahmoud@elkhouly.net

*Abstract*
*We are taught from a young age that plagiarism (copying other's work) is wrong. However, the problem of Illegal copies of multimedia data is exacerbated by the widespread availability of circumvention devices, which enable people to make infringing copies of multimedia data. Recently, Joint Video Compression and Encryption (JVCE) has gained increased attention to reduce the computational complexity of video compression, as well as provide encryption of multimedia data. In this paper, a novel protection method for multimedia data (ECAKP) is proposed. It combines encryption process and compression with authenticating process. The method had been implemented and the results are discussed in detail.*

## 1. INTRODUCTION

According to the Business Software Alliance about 35% of the global software market, worth $141 Billion, is counterfeit. Most of the counterfeit software is distributed in the form of a compact disc (CD) or a digital video disc (DVD) which is easily copied and sold in street corners all around the world but mostly in developing countries [1]. The current e-learning global market size is over $20 billion – grown ten-fold since 2000 [2]. E-learning market is expected to surpass $52.6 billion by 2010. Online tutoring is a $4 billion industry and is growing at a rate of 10% - 15% per annum. Today, Universities become producers of electronic contents such as copyrighted videos, CD-ROMs, websites and courseware [3]. Also, some multimedia data in politics, economics or militaries are necessary to be protected.

In the past decade, some multimedia encryption algorithms have been reported. According to the relationship between encryption process and compression process, these algorithms can be classified into two types: the first one encrypts the compressed data completely or partially; the second one encrypts the data during compression [4].

### 1.1 Multimedia Compression Basics

Video or multimedia compression refers to reducing the quantity of data used to represent digital video data. It allows transmission of multimedia over bandwidth constrained communication channels. For example- a standard video monitor displays a frame usually with the resolution of 800×600 pixels. For a color image, a pixel is represented by 3 bytes of data (one for Red, Blue and Green respectively). Thus, a one hour video at 30 frames per second will require 144 GB of space in hard disk and is impossible to transmit it over any practical communication channel. It is compressed to around 500- 600 MB by the use of MPEG-2 [5] compression format. Tseng, et al. [6] gave a good summary of recent advances in multimedia compression [7].

## 1.2 Multimedia Encryption Basics

Multimedia encryption protects multimedia data by transforming it into an unrecognized format. Only authorized users with the correct security keys are able to decode and view protected data. Images or videos can be encrypted either partially or fully using various technologies in the spatial or frequency domains. Image/video encryption in the frequency domain is often embedded in the compression process, which is mainly based on the Discrete Cosine Transform (DCT) [8] or Discrete Wavelet Transform (DWT) [9]. Image/video encryption in the spatial domain changes pixel locations and/or values using different techniques. Data Encryption Standard (DES) (http://csrc.nist.gov/publications/fips/fips46-3/fips46-3.pdf) and Advanced Encryption Standard (AES) [10] are two examples of this method. However, they do have high computation costs [11]. Other techniques include chaos theory [12] and recursive sequences [13]. Nevertheless, due to their lack of security keys or the small key space, these approaches often involve either high computation costs or provide low levels of security [14].

## 1.3 Copy-protected Cds

Since 2002, millions of copy-protected CDs have been released. One common method that has been used to protect CDs from being copied includes adding data to the CD copy that makes it unreadable to copiers; a second method includes altering the way the files are listed in the table of contents so that the CD can't be read by a CD-ROM.
No matter how frequently the multimedia industries attempt to build the ultimate CD copy blocking system, enterprising people will find ways around them. For example, a university student found if you hold down the shift key while the CD begins to load onto your machine the Microsoft Windows (AutoRun) feature prevents the anti-piracy software from loading. This doesn't delete the program but bypasses the copy protection installation, making the music available for copying. Once such a discovery is made public on the Internet, the rest is history (http://cd-burning-software-review.toptenreviews.com/cd-encryption.html).

## 1.4 DRM

Short for digital rights management, a system for protecting the copyrights of data circulated via the Internet or other digital media by enabling secure distribution and/or disabling illegal distribution of the data. Typically, a DRM system protects intellectual property by either encrypting the data so that it can only be accessed by authorized users or marking the content with a digital watermark or similar method so that the content can not be freely distributed (http://www.webopedia.com/TERM/D/DRM.html).

## 1.5 Product activation

Product activation is widely used by software vendors to protect their applications and enforce license agreements. A key concern for software vendors is ensuring users don't just give the software to unlicensed friends and colleagues, or even post it on the web for anyone to download. Older approaches for license enforcement include dongle-based licensing and key-file-based licensing. A dongle is a hardware device that plugs into the user's computer; when the application runs it checks for the presence of the dongle and will run only if it finds it. Dongles do therefore allow the user to move their license around, but only by physically relocating the dongle. With key-file-based licensing, the license limits and node-locking parameters are encrypted in a file, which is sent to the user and read by the application each time it runs. These approaches have a number of disadvantages. Dongles require the distribution of the hardware, with all that entails in material cost, shipping cost, delivery times and management by the vendor. Product activation systems therefore meet the software vendors' need to protect against piracy, offer a range of license models, and automate operations, but remove many of the inconveniences and costs of older license management systems [15].

## 2. PROPOSED ECAKP

The primary goals of ECAKP are to protect multimedia files from being copied illegally, and to do it with on-shelf tools available for most people. ECAKP has four phases; first it encrypts the original multimedia data to unrecognized format and compresses it. At second phase, it collects some non personal data from the customer's computer set during his/her installation the program. Third phase is authentication one, in which, comparison between collected data and the list of authorized customers who purchased the original copies took place, then it decompresses the encrypted multimedia file. Final phase has two steps: (a) it kills recording programs running on the computer, (b) disconnects internet connection (science some portable programs can record the computer's screen, and therefore can get illegal copy). For notation see Table 1. Figure 1 shows the processes of ECAKP .

Table 1. Notations used in ECAKP

| | |
|---|---|
| P,U,PC,S | Producer, user, user PC, and the server, respectively. |
| V, M, I | Original multimedia file, encrypted multimedia file, and install file, respectively. |
| $R_E$, F | Random number generated by P, license file, respectively. |
| $ID_M$, $ID_{PC}$ | Video ID and PC ID, respectively. |
| F(.) | A cryptographically secure hash function. |
| V(.) | A visualization function that decrypt any bit stream. |
| C, D, K, Pl | Collect, decrypt, kill and play processes, respectively. |

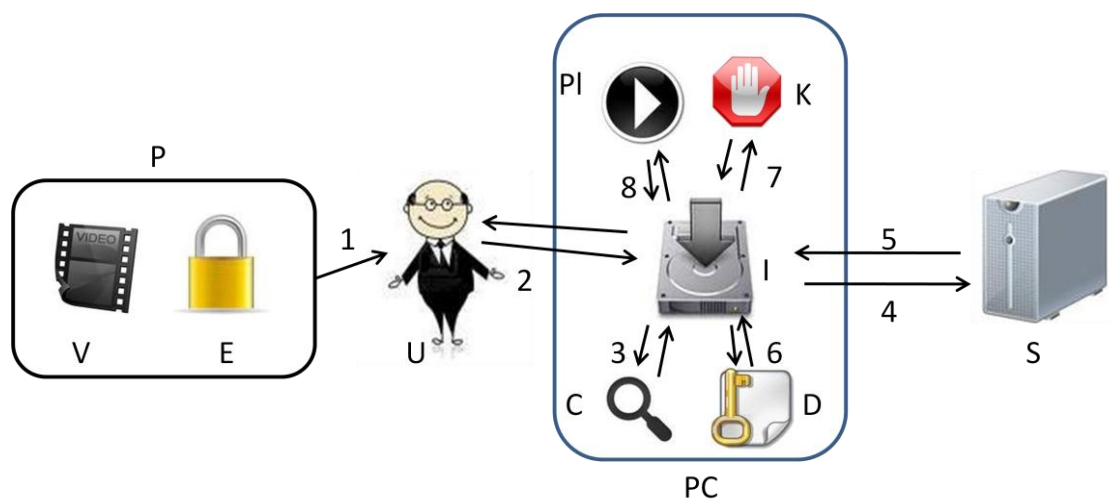

Fig. 1 ECAKP steps

1. P encrypts V using a random nonce $R_E$ produce M with $ID_M$.

$$U \leftarrow P: M_{ID_M} \qquad (1)$$

   Here is a brief description of the header content, some of them are omitted:
   First byte : ID1 = 0×1F
   Second byte: ID2 = 0×8B
   (These are "mathematical key" that describe / identify GZIP compression)
   Compression method:
   0×0-7 reserved
   0×08 DEFLATE (that uses a combination of the LZ77 algorithm and Huffman coding) [16].

2. U installs I on the PC, then I prompts the user to input $ID_M$ and the user's e-mail.

3. I invokes C , which uses a snapshot of the end customer's PC to create a system identity. Activation uniquely associates the software license with this system identity - the license can only be unlocked on the system with the associated identity. This hardware fingerprinting is a sophisticated technique for tying an application to a specific computer. As a result, users who update their hardware or operating systems won't need to obtain a new activation code and re-activate their license.

$$I \leftarrow C: ID_{PC} \qquad (2)$$

4. I sends both program's ID and PC's ID from (2) through the following message to S.

$$S \leftarrow I: ID_M, ID_{PC} \qquad (3)$$

5. At S, there are several pre-defined policies that can use to manage licenses. This is a per-product setting.
   - *Monitor only mode.* In this mode S will not block any activation requests. It will simply log the requests and record the statistics.
   - *Massive fraud prevention only.* This mode will only refuse to give the activation code after a fair amount of activations is made from different computers. Such registration number will then be marked as stolen and invalid.
   - License terms enforcement – fair use. In this mode S would only allow a fair number of complimentary activations per purchased copy. A 'fair' use for a general purpose program would be single additional activation per copy (say, business & home, or stationary & laptop computers).
   - *Strict license terms enforcement.* An activation code will be issued for a single PC only; re-activation will succeed on the same PC but fail on a different one.
   
   Then, S generates the license file $F_{ID_{PC}}$ and forwards it to I

$$I \leftarrow S: F_{ID_{PC}} \qquad (4)$$

6. I decrypts $M_{ID_M}$ from (1) using $F_{ID_{PC}}$ to recover V($R_E$). Then I displays success or failure to U. In case of success U can play V,

$$I \leftarrow V(M_{ID_M}, F_{ID_{PC}}) \qquad (5)$$

The following code shows decryption process.

```
FileStream source1 = File.OpenRead(Form1.CdPath +
dataGridView1.SelectedRows[0].Cells[0].Value.ToString());
GZipStream gZip1 = new GZipStream(source1,
CompressionMode.Decompress);
FileStream Decompressed = new
FileStream(Environment.GetFolderPath(Environment.SpecialFolder.
System) + @"/Temp"+x,FileMode.Create);
int b1 = gZip1.ReadByte();
while (b1 != -1)
{
Decompressed.WriteByte((byte)b1);
b1 = gZip1.ReadByte();
}
gZip1.Close();
Decompressed.Close();
source1.Close();
gZip1.Dispose(); Decompressed.Dispose(); source1.Dispose();
LessonPath =
Environment.GetFolderPath(Environment.SpecialFolder.System) +
@"/Temp"+x;
Frm_Player frm = new Frm_Player();
frm.ShowDialog();
frm.Dispose();
File.Delete(LessonPath);
x++;
```

7. I invokes K. To ensure multimedia contents will not be recorded during lunching the software. K prevents the following programs to work during playing:
   - Virtual CD programs
   - Internet connection (except when activate the program)
   - Video recording programs

   The following code is used to list some processes that allowed running during playing the multimedia.

```
void ClosePrograms()
{aa.Add("svchost"); aa.Add("lsass"); aa.Add("services");
aa.Add("winlogon"); aa.Add("csrss"); aa.Add("smss");
 aa.Add("System"); aa.Add("notepad"); aa.Add("Idle");
 aa.Add("spoolsv"); aa.Add("alg"); aa.Add("WINWORD");
 aa.Add("AcroRd32"); aa.Add("explorer"); aa.Add("devenv");
 aa.Add("sqlservr");
 foreach (Process var in Process.GetProcesses())
  {if (aa.Contains(var.ProcessName) == false)
  {if (var.ProcessName !=
Process.GetCurrentProcess().ProcessName)
   { try   {   var.Kill();}
     Catch {continue;}}}}
 Counter = Process.GetProcesses().Length;}
```

8. I plays D.

## 3. CASE STUDY

The programming language course at Faculty of Science, Helwan University is presented through one hour lectures and one hour lab per week for one semester. The lectures are conducted with the entire enrollment in a large lecture hall. Laboratory sections are divided into 18 groups of 30 students each. Attendance at the laboratory sections is highly recommended, but not required. This "open" laboratory policy permitted students to use their home computers for the required assignments, as well as the computers in the programming laboratory. The course is required for all second year students (total number of students is 540 students).

### 3.1 CD instead of textbook

The course materials were available to students in the form of compact disks (CD's) that the students could purchase, like a textbook from the bookstore. The CD contains the encryptions files resulted from first phase, and the setup file. It could be used on students' own personal computers or on the personal computers provided by the University in the laboratory. The topics covered by the CD were those found in a typical college entry level programming courses, e.g. standard output, standard input, auto increment and decrement, If-else statement, loops and arrays. Navigation is managed with the menu bar and allows users to navigate directly to any chapter or subsection in the CD. Using a combination of video and audio, the tutorials explained many of the basic principles of programming language which would be difficult with conventional textbooks or lectures notes alone. A series of interactive example problems, with solutions, that parallel the course lectures and assignments were part of the CD. These examples provided step-by-step solutions to traditional programming language problems, explaining at each step the proper procedure to reach the solution.

### 3.2 Data gathering

Every year, students used to buy about 400 textbooks for programming course. In 2010 it was the first time to introduce CD with both video and text contents for this course. The purchasing was slow at the beginning due to the thought of students that they can activated the illegal copies they got, however, after they trials failed, the total number of purchasing was 190 CDs by the end of the course. The total number of successful activation was 182, and 8 students never try to activate their copies!.
At the end of the course, students bought CDs were asked some questions as:
   a) Do you have access to computer and Internet?
      40.3% reported having access to a computer and Internet at home. Of these, 38% had broadband access. 50.7% reported having access to a computer and Internet at University campus, while 9% reported having access to a computer and Internet at Net café .
   b) Do you get trouble during activation phase?
      Only 3% reported having trouble. By analysis we obtained that the troubles can be classified as:
      - Try to write a password in non-sensitive manner.
      - No updated operating system installed.
   c) What would you describe as your primary reasons for buying the course' CD?
      The top six reasons in which the students reported are listed in Table 2.

Teacher is asked to provide answer to the open-ended question:
"What would you describe as your primary goals for using course's CD instead of course's textbook?" Teacher points out the following goals:
   a) Intellectual property: science students used to copy his textbook, and they cannot copy his protected CD.
   b) Cost benefit: science producing one protected CD is much cheaper than printing one textbook.

c) Improve learning by providing video materials as well as text materials.
d) Provokes university to continue to prioritize e-learning.

Table 2. Reasons for buying course' CD

|  | N | % |
|---|---|---|
| Access course video materials | 172 | 94.5 |
| Test out their knowledge and receive feedback using the quiz facility | 46 | 25.3 |
| Time and place flexibility | 35 | 19.2 |
| Access course materials prior to lectures | 12 | 6.5 |
| Qualitatively different learning opportunities | 11 | 6.0 |
| Locate other learning resources via links provided | 6 | 3.2 |

## 4. CONCLUSION

For multimedia systems, user's authentication is most essential in association with the access control of the multimedia system. At activation process over the Internet, additional data encryptions are also adopted in protection of the multimedia. In this paper, ECAKP is presented. According to ECAKP, the encryption scheme combined with compression process is not only practical but also secure by choosing a suitable encryption algorithm and activation over the internet. ECAKP deters software piracy and offers cost efficiency for producer – avoiding the cost of providing users with expensive protection software or hardware tokens (as well as the token maintenance cost).Case study showed how ECAKP is effectively used to protect multimedia CD contents, and therefore, protect intellectual property of the author. ECAKP processes are simple, less cost than commercial encryption software (or even has no cost, if the beneficiary will implement it by himself). In the future work, selected parts of the multimedia data will only be encrypted during compression process to speed up the runtime decryption process.

## ACKNOWLEDGMENTS

This study was supported by Quality System company (www.quality-sys.com). I would like to acknowledge the support of Mr. Abd Allah Abd El Aleem who helped in developing the various activities to create the CDs for this course.

## REFERENCES

[1] Ghaith H., Aykutlu D. & Berk S (2009). CDs Have Fingerprints Too. Proceedings of the 11th International Workshop on Cryptographic Hardware and Embedded Systems, Lausanne, Switzerland, Pp.348 – 362, ISBN:978-3-642-04137-2.

[2] Jayanthi C., Reema Jose (2008), Investing in e-learning proves to be lucrative
http://www.financialexpress.com/news/investing-in-elearning-proves-to-be-lucrative/314400/0

[3] Nisachol Channgern & Settapong Malisuwan, Proceedings of the Second International Conference on eLearning for Knowledge-Based Society, August 4-7, 2005, Bangkok, Thailand.

[4] Dengpan Ye & Shiguo Lian, (2009). "Novel encryption model for multimedia data", Journal of Systems Engineering and Electronics, Vol. 20, No. 5, pp. 1081-1085.

[5] Barry G. Haskell, A. P. and Netravali, A. N. (1999). Digital Video: An Introduction to MPEG-2.Springer.

[6] Tseng, P., Chang, Y., Huang, Y., Fang, H., Huang, C., and Chen, L. (2005). Advances in Hardware Architectures for Image and Video Coding - A Survey. Proc. IEEE, 93(1):184–197.

[7] Amit Pande (2010). Algorithms and architectures for secure embedded multimedia systems, PhD thesis, Iowa State University, Ames, Iowa.

[8] Ci Wang, Hong-Bin Yu, and Meng Zheng (2003). A DCT-based MPEG-2 transparent scrambling algorithm," IEEE Transactions on Consumer Electronics, vol. 49, no.4, pp. 1208-1213.

[9] Guo-Sheng Gu and Guo-Qiang Han (2006). "The Application of Chaos and DWT in Image Scrambling," in 2006 International Conference on Machine Learning and Cybernetics, pp. 3729-3733.

[10] Joan Daemen and Vincent Rijmen (2000). "The Block Cipher Rijndael," in Proceedings of the The International Conference on Smart Card Research and Applications: Springer-Verlag.

[11] M. Grangetto, E. Magli, and G. Olmo (2006). "Multimedia Selective Encryption by Means of Randomized Arithmetic Coding," IEEE Transactions on Multimedia, vol. 8, no. 5, pp. 905-917.

[12] Mohamed Amin and Ahmed A. Abd El-Latif (2010). "Efficient modified RC5 based on chaos adapted to image encryption," Journal of Electronic Imaging, vol. 19, no. 1, pp. 013012-10.

[13] Rong-Jian Chen and Jui-Lin Lai (2007). "Image security system using recursive cellular automata substitution," Pattern Recognition, vol. 40, no. 5, pp. 1621-1631.

[14] Yicong Zhou, (2010). Multimedia Security System for Security and Medical Applications, PhD thesis, TUFTS University.

[15] http://www.articlesbase.com/programming-articles/why-product-activation-for-software-is-becoming-widespread-913435.html

[16] Thomas H. Cormen, Charles E. Leiserson, Ronald L. Rivest, and Clifford Stein (2001). Introduction to Algorithms, Second Edition. MIT Press and McGraw-Hill, ISBN 0-262-03293-7. Section 16.3, pp. 385–392.

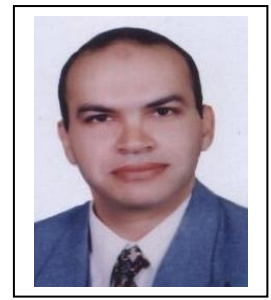

**Assistant Professor Mahmoud Mohamed El-Khouly** received his BSc degree in Mathematics from Helwan University in 1983, his first Master Degree from the same University / Egypt (1994), his second Master Degree from Cairo University in computer sciences/ Egypt (1995), and his Doctorate of Philosophy from Saitama University in computer sciences/ Japan (2000). Assistant Professor El-Khouly helds different academic positions at Temple University Japan (TUJ) (2001). Lecturer in Helwan University (2001-2005). Visiting Professor at Qatar University (2003-2004). He acted as Cultural Attache' in Egyptian Embassy – London (2005-2008). He is now working as assistant professor at Faculty of Computer and Information, Helwan University Egypt.